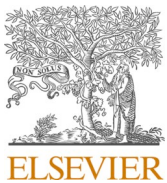
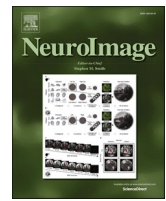

# Neural modulation enhancement using connectivity-based EEG neurofeedback with simultaneous fMRI for emotion regulation

Amin Dehghani [a,b,*], Hamid Soltanian-Zadeh [a,c,d], Gholam-Ali Hossein-Zadeh [a,c]

[a] *School of Electrical and Computer Engineering, University of Tehran, Tehran, Iran*
[b] *Department of Psychological and Brain Sciences, Dartmouth College, Hanover, NH, USA*
[c] *School of Cognitive Sciences, Institute for Research in Fundamental Sciences (IPM), Tehran, Iran*
[d] *Departments of Radiology and Research Administration, Henry Ford Health System, Detroit, MI, USA*

ABSTRACT

Emotion regulation plays a key role in human behavior and overall well-being. Neurofeedback is a non-invasive self-brain training technique used for emotion regulation to enhance brain function and treatment of mental disorders through behavioral changes. Previous neurofeedback research often focused on using activity from a single brain region as measured by fMRI or power from one or two EEG electrodes. In a new study, we employed connectivity-based EEG neurofeedback through recalling positive autobiographical memories and simultaneous fMRI to upregulate positive emotion. In our novel approach, the feedback was determined by the coherence of EEG electrodes rather than the power of one or two electrodes. We compared the efficiency of this connectivity-based neurofeedback to traditional activity-based neurofeedback through multiple experiments. The results showed that connectivity-based neurofeedback effectively improved BOLD signal change and connectivity in key emotion regulation regions such as the amygdala, thalamus, and insula, and increased EEG frontal asymmetry, which is a biomarker for emotion regulation and treatment of mental disorders such as PTSD, anxiety, and depression and coherence among EEG channels. The psychometric evaluations conducted both before and after the neurofeedback experiments revealed that participants demonstrated improvements in enhancing positive emotions and reducing negative emotions when utilizing connectivity-based neurofeedback, as compared to traditional activity-based and sham neurofeedback approaches. These findings suggest that connectivity-based neurofeedback may be a superior method for regulating emotions and could be a useful alternative therapy for mental disorders, providing individuals with greater control over their brain and mental functions.

## 1. Introduction

Emotion regulation is an intrinsic or extrinsic process of managing and modifying emotional experiences, which can include enhancing, inhibiting, or maintaining emotional responses through changes in behavior, feelings, or physiological reactions (Gross, 2015; Thompson, 2019). Emotion regulation plays a key role in behavior and one's life quality. Several strategies including situation selection, situation modification, attention deployment, cognitive change, and response modulation have been proposed for emotion regulation (Gross, 1998).

Neurofeedback (NF) is a noninvasive self-brain training technique used for emotion regulation to enhance brain function or treatment of mental disorders leading to behavioral changes (Brewer et al., 2011). By providing individuals with real-time feedback on their brain activity, neurofeedback allows them to gain voluntary control over their neural processes associated with emotions. This technique operates on the principle of neuroplasticity, which suggests that the brain can adapt and reorganize its functioning based on experience and feedback (Loriette et al., 2021). Through neurofeedback training, individuals can learn to modulate their brain activity and achieve a more desirable emotional state. Unlike invasive procedures that involve surgical interventions or pharmacological approaches that introduce external substances into the body, neurofeedback relies on the use of different devices and techniques to measure brain activity. This makes it a safe and well-tolerated method, minimizing potential risks and side effects associated with more invasive techniques. Furthermore, neurofeedback has shown promise in addressing a wide range of emotional dysregulation issues. Research studies have explored its application in various clinical conditions, such as anxiety disorders, depression, post-traumatic stress disorder (PTSD), and attention deficit hyperactivity disorder (ADHD). By training individuals to self-regulate their brain activity patterns, neurofeedback has the potential to reduce symptoms, enhance emotional well-being, and improve overall mental health outcomes (Ahrweiler et al., 2022; Ciccarelli et al., 2023; Louthrenoo et al., 2022; Moreno-García et al., 2022; Russo et al., 2022; Shaw et al., 2023; Taylor et al., 2022; Zhao et al., 2023).

Electroencephalography (EEG) and functional magnetic resonance imaging (fMRI) are two neuroimaging techniques commonly used in neurofeedback research. EEG neurofeedback has several medical and non-medical applications (Bolea, 2010; Coben et al., 2010; Gapen et al., 2016; Quaedflieg et al., 2016; Zuberer et al., 2018). EEG frontal asymmetry has been utilized in numerous studies to investigate emotion regulation, particularly based on the approach-withdrawal model (Davidson et al., 1990; Mennella et al., 2017; Peeters et al., 2014; Wang et al., 2019). According to this model, negative emotions or withdrawal-related affect (e.g., fear, sadness, and disgust) are linked to increased activity in the right hemisphere. Conversely, heightened activity in the left hemisphere is associated with emotional states like joy or anger (Quaedflieg et al., 2016). Multiple prior studies have highlighted the potential of EEG frontal asymmetry as an indicator of distinct emotional states and as a biomarker for conditions such as PTSD, anxiety, and depression (Coan and Allen, 2004; Deng et al., 2023; Koller-Schlaud et al., 2020; López-Castro et al., 2021; Meyer et al., 2018; Meza-Cervera et al., 2023; Spironelli et al., 2021; Stead et al., 2023). It was hypothesized that individuals have the ability to modulate positive and negative affect through EEG frontal asymmetry neurofeedback (Peeters et al., 2014b). The findings indicated that increased activity in the left frontal hemisphere was associated with positive affect, while increased activity in the right frontal hemisphere was linked to negative affect. Additionally, decreased activity in the right frontal hemisphere was associated with positive affect, while decreased activity in the left frontal hemisphere was linked to negative affect. Previous studies have employed simultaneous recording of EEG and fMRI during emotion regulation, often using EEG asymmetry in the frontal region alone or in combination with fMRI as a form of neurofeedback (Dehghani et al., 2022, 2020; Mosayebi et al., 2022; Zotev et al., 2016, 2014). Previous neurofeedback studies have mainly been restricted to utilizing the activity of a single brain region as measured by fMRI or the power of one or two electrodes in EEG in order to calculate neurofeedback (Arpaia et al., 2022; Cavazza et al., 2014; Dehghani et al., 2022; Herwig et al., 2019; Marxen et al., 2016; Mennella et al., 2017; Misaki et al., 2018; Paret et al., 2018; Wang et al., 2019; Zhu et al., 2019; Zotev et al., 2018, 2014).

According to previous studies (Kim et al., 2015; Sulzer et al., 2013; Zotev et al., 2011), feedback based on network or connectivity may result in better/higher regulation through feedback. In this study, we hypothesize that utilizing neurofeedback to regulate emotions through coherence of EEG electrode leads to greater modulation of brain function and activity compared to traditional frontal asymmetry neurofeedback. Additionally, we propose that this approach offers enhanced volitional control over brain and mental function. To investigate this, we administer neurofeedback based on the coherence of EEG electrodes, building upon the findings from our prior study (Dehghani et al., 2021). The findings of this study highlight the efficacy of connectivity-based neurofeedback in modulating EEG frontal asymmetry, EEG coherence, fMRI BOLD signal/connectivity, and psychometric test outcomes, surpassing the effects observed with traditional EEG frontal asymmetry neurofeedback and a sham control group.

## 2. Methods

### 2.1. Task design

The research protocol was approved by the ethics committees of the Iran University of Medical Sciences, Tehran, Iran. This study comprises three experimental groups (groups 1, 2, and 3). The data for groups 2 and 3 is derived from pre-existing data utilized in previous studies (Dehghani et al., 2022, 2020; Mosayebi et al., 2022). Ten healthy subjects (age 26.5 ± 3.6 years, all-male, white) for connectivity-based EEG neurofeedback (experimental group 1) participated in this study. The results of this study were compared with those of our previous study that used EEG frontal asymmetry neurofeedback as the experimental group (experimental group 2 including 20 healthy subjects and age 26.7 ± 3.6 years, all male, white), as well as with a sham control group (experimental group 3 including 15 healthy subjects and age 27 ± 3.8 years, all male, white). Participants in the control group were provided (without their knowledge) with sham EEG neurofeedback. During the quality control of data, no dropouts were detected. However, prior to the commencement of the experiments, three individuals (two from the experimental group 2 and one from the experimental group 3) dropped out due to technical issues for MR scanner. Participants in experimental group 2 received neurofeedback based on EEG frontal asymmetry in the alpha frequency band, while those in experimental group 1 received neurofeedback based on the coherence of EEG electrodes. Participants in the control group received sham feedback without being aware that it was not real feedback. The study's exclusion criteria encompassed individuals with a current or prior history of significant psychiatric or neurological disorders, substance abuse within the past year, previous brain surgery, and issues related to undergoing MRI scans. All participants were right-handed and had either normal vision or vision corrected to normal. Additionally, prior to the experiment, each participant completed two psychometric tests including Beck's Depression Inventory (Craven et al., 1988) and the General Health Questionnaire - 28 (GHQ-28) (Nazifi et al., 2013). The mean scores for Beck's Depression Inventory and GHQ-28 were 6.6 ± 3.2 and 2.1 ± 2.1, respectively. Based on these scores, the participants exhibited normal levels of depression according to Beck's Depression Inventory and were classified as non-psychiatric based on the GHQ-28 test result. All participants were examined by a resident physician before the experiment and filled out the consent form.

The study used a paradigm of retrieving positive autobiographical memories, as previously described in (Dehghani et al., 2022, 2020). Before the experiment, we had an interview with each participant and we asked them to tell several positive autobiographical memories, and individualized images were chosen based on each memory. The experiment consisted of 10 runs, each comprising a rest block (20 sec), a view block (40 sec) and an upregulation block (60 sec). During the rest block, subjects were instructed to relax with open eyes, while no images were displayed. During the view block, participants were presented with two pictures depicting their positive autobiographical memories that were mentioned during the interview. They were instructed to simply observe the images without actively trying to recall the associated memories. In the upregulation block, participants were presented with similar images as in the previous view block. They were instructed to increase the height of neurofeedback bar by recalling and retrieving related autobiographical memories. Before the neurofeedback session, participants were given a sample run of the paradigm without viewing individual pictures.

The neurofeedback used in experimental group 2 was based on the approach-withdrawal hypothesis (Davidson, 1998), which involves measuring the difference in EEG power between the right and left hemispheres in the alpha frequency band in 2-second time windows with 50% overlap. The neurofeedback for experimental group 1, on the other hand, was based on the coherence of EEG channels F4 and F3 within the alpha frequency band. Further details regarding this approach are discussed in subsequent sections. A recent study has shown that the coherence between the F4 and F3 electrodes in the alpha frequency band during emotion regulation blocks increases significantly compared to other blocks of the experiment (Dehghani et al., 2021). Therefore, the change in coherence between these two electrodes was measured in 2-second time windows with 50% overlap and presented to participants as feedback. The experimental protocol for a single run is shown in Fig. 1.

*2.2. Data acquisition*

The MRI scans were obtained using a 3 Tesla Prisma scanner (Siemens, Erlangen, Germany) at the National Brain Mapping Lab (NBML) in Tehran, Iran. Structural images were captured using a T1-weighted MPRAGE pulse sequence (TI = 1100 msec, TR = 1810 msec, TE = 3.47 msec, flip angle = 90°, and voxel size = 1×1×1 mm), and functional data were obtained using a T2*-weighted gradient-echo, echo-planar (EPI) pulse sequence (TR = 2000 msec, TE = 30 msec, flip angle = 90°, matrix size = 64×64×30, and voxel size = 3.8×3.8×4 mm). Over the course of 10 experimental runs, 650 vol images were acquired. EEG data were recorded at 5k samples/sec using a 64 channel MR-compatible EEG system (Brain Products, München, Germany) in accordance with the 10–20 system with AFz serving as the ground electrode and FCz as the reference electrode. Electrode impedance was kept below 5 K Ohms. The task was displayed via a coil-mounted display, which was controlled by the Psychtoolbox program and allowed participants to view each block of the paradigm during the neurofeedback experiment.

## 3. Data analysis

*3.1. EEG and fMRI data analysis*

Due to practical limitations, neurofeedback was only based on the EEG signal. For the online analysis, the MRI gradient and ballistocardiogram artifacts were removed using a moving average algorithm through BrainVision RecView software (Brain Products GmbH). The average head movement was less than 0.38 mm for all participants, therefore removing MRI gradient and ballistocardiogram artifacts in real-time was effective as well as counterpart methods (Moosmann et al., 2009; Niazy et al., 2005).

Quality control procedures for both online and offline EEG data assessment were described in previous studies (Dehghani et al., 2022, 2020). In connectivity-based neurofeedback, the denoised data was downsampled to 250 samples/second. During the "Upregulation" block, the coherence between channels F3 and F4 was calculated using a 2-second moving window and presented as a bar during the upregulation blocks, relative to the baseline (coherence between channels F3 and F4 in previous "View" block). The neurofeedback values were updated every 1 second.

The offline preprocessing of both EEG and fMRI data were described in full detail in our previous studies (Dehghani et al., 2022, 2020). In summary, the pre-processing of a single subject's fMRI data was performed using the FSL software package (S.m. et al., 2005; Smith et al., 2004; Woolrich et al., 2009) and included removing the first 10 vol of fMRI data, slice-timing correction, motion correction, temporal high-pass filtering (cut-off = 0.005 Hz), spatial smoothing using an 8 mm full-width at half-maximum Gaussian kernel, and standard GLM analysis with three regressors for rest, view, and upregulation blocks. Additionally, the analysis included regressors for motion (24 regressors), CSF, and global mean signal confounds. Statistic images were thresholded using clusters determined by Z>2.5 and a cluster significance threshold of P=0.01 and was corrected for multiple comparison using Gaussian Random Field (GRF) (Smith et al., 2004; Woolrich et al., 2009). The mean BOLD signal of activated regions in the preprocessed fMRI data, registered to the Montreal Neurological Institute (MNI) atlas, was calculated for different contrasts of upregulation versus view and rest ($\frac{\text{mean(Upregualtion)} - \text{mean(View)}}{\text{mean(View)}}$ or $\frac{\text{mean(Upregualtion)} - \text{mean(Rest)}}{\text{mean(Rest)}}$) in the 118 anatomical masks from "WFU_PickAtlas" and FSL (Desikan et al., 2006; Gorgolewski et al., 2015; Tzourio-Mazoyer et al., 2002). In addition, Pearson's correlation was employed to estimate the functional connectivity within different blocks of the experiment. For this purpose, anatomical masks of distinct brain regions are utilized to extract the average BOLD signals from the activated voxels within those regions. Subsequently, the functional connectivity between these regions was calculated.

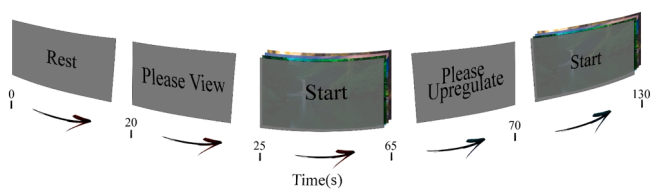

**Fig. 1.** The diagram of the neurofeedback paradigm consisted of rest, view, and upregulation blocks (Dehghani et al., 2022).

The offline analysis of the EEG data for each participant included two main steps to remove artifacts and was performed using the FMRIB plug-in as a Matlab toolbox. The first step was removing the MRI gradient and the second step was removing the ballistocardiogram artifact. The details of these two steps including 4 steps (adjusting slice-timing triggers, mean average template subtraction, optimal basis sets, and adaptive noise cancelation) for MR artifact and 2 steps (QRS detection and optimal basis sets) for BCG artifact were discussed in (Dehghani et al., 2022; Niazy et al., 2005). After removing the MR and BCG artifacts, the EEG data were down-sampled to 250 samples/sec and low-pass filtered at 100 Hz. The fMRI slice selection frequency and its harmonics were removed by bandpass filtering. Independent component analysis (ICA) was then applied over the entire EEG data (excluding noisy and motion-affected intervals). Independent components (ICs) corresponding to different artifacts, such as eye blink, head movement, ballistocardiogram, or BCG residual were identified and removed using time course, spectral, topographic map and kurtosis.

*3.2. Psychometric testing*

To assess changes in mood state, participants completed a short Persian version of the Profile of Mood States (POMS), the complete Persian version of the Positive-Negative Affect Scale (PANAS), and the Persian version of the Beck Anxiety Inventory (BAI) questionnaire before and after the neurofeedback test (Ghassamia et al., 2013; Khesht-Masjedi et al., 2015)

## 4. Results

Fig. 2. Presents the activation maps for different experimental groups including activity, connectivity-based and sham neurofeedback, and for connectivity-based vs. activity-based neurofeedback, activity-based vs. sham neurofeedback, and connectivity-based vs. sham neurofeedback groups. These maps highlight various brain regions within the frontal, temporal, occipital, and limbic systems.

Table 1. Summarizes the signal changes of these activated brain regions in various experiments, including comparisons of upregulation versus view and rest, as well as previous studies (Repeated Measures ANOVA with Upregulation, View, and Rest as independent variable for every one of the 118 ROIs used in this study, FDR-corrected for multiple comparison; q = 0.05) and the corresponding t-value.

The signal change percentage in the connectivity-based neurofeedback experiment was found to be higher than that of the activity-based neurofeedback experiment and previous studies (Bado et al., 2014; Burianova et al., 2010; Ino et al., 2011; Johnston et al., 2011, 2010; Kim and Hamann, 2007; Lempert et al., 2017; Li et al., 2016; Pelletier et al., 2003; Zotev et al., 2014, 2016). These higher percentage changes are related to the effectiveness of the connectivity-based neurofeedback training used in this study. As also seen in Table 1, the signal change and activity peak in most brain regions, including deep brain regions such as the amygdala, insula, and thalamus, as well as frontal/prefrontal regions like the orbitofrontal cortex or dorsolateral prefrontal cortex, were found to be higher in the connectivity-based neurofeedback compared to the activity-based neurofeedback.

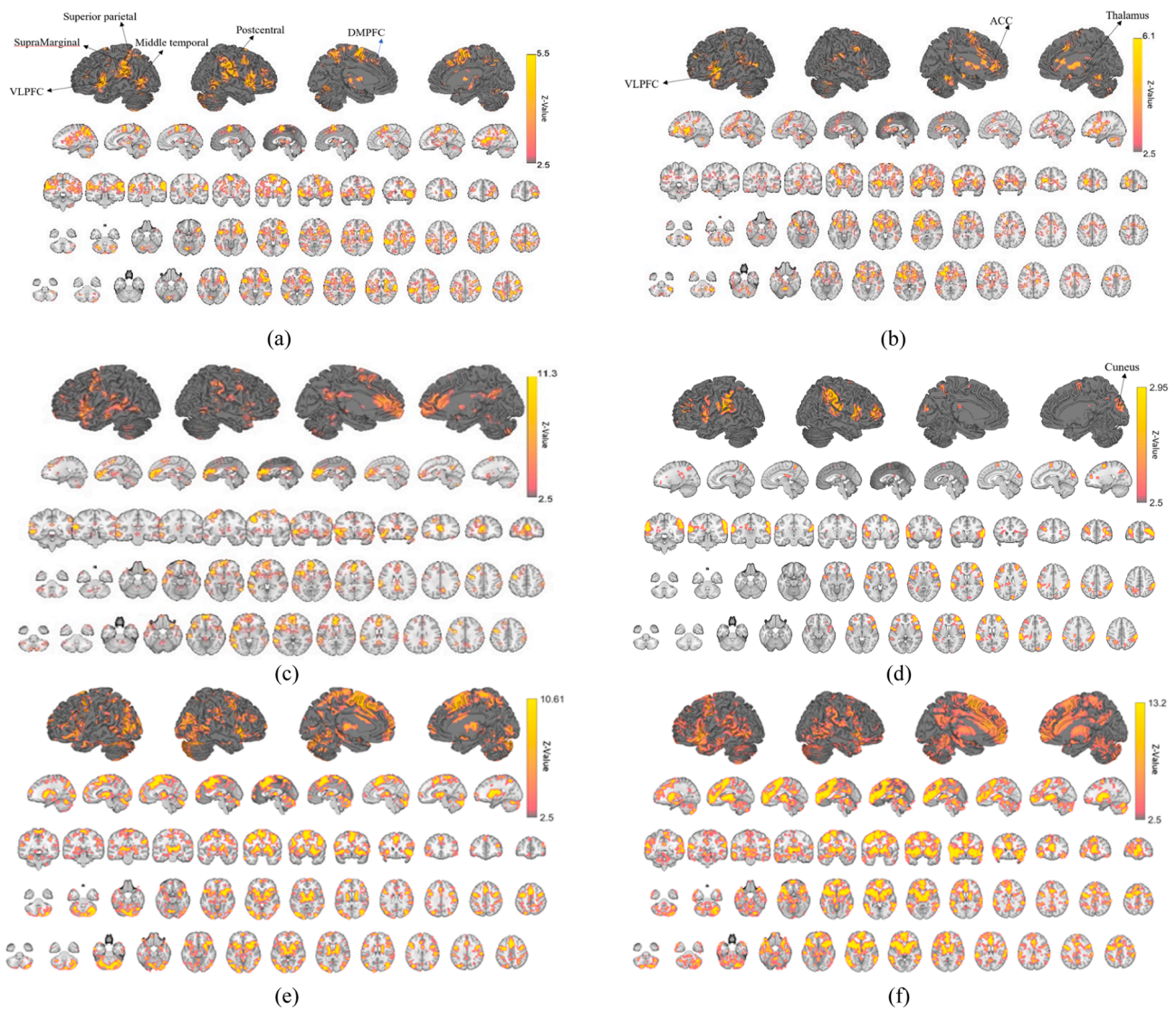

**Fig. 2.** Activation map (Z-Value) for upregulation versus view for (a) activity-based neurofeedback (b) connectivity-based neurofeedback (c) connectivity vs. activity-based neurofeedback (d) sham neurofeedback (e) activity-based vs. sham neurofeedback (f) connectivity-based vs. sham neurofeedback.

In our previous study using activity-based neurofeedback (Dehghani et al., 2022), functional connectivity was calculated based on the Pearson correlation of BOLD signal time series among thirty-eight brain regions (left/right amygdala, thalamus, insula, dorsomdical prefrontal cortex (DMPFC), caudate, cuneus, hippocampus, posterior cingulate cortex, orbitofrontal cortex (OFC), middle temporal gyrus, lingual gyrus, ventral striatum, dorsolateral prefrontal cortex (DLPFC), ventrolateral prefrontal cortex (VLPFC), superior parietal, inferior parietal, supramarginal, postcentral, and anterior cingulate cortex (ACC)). The comparison of Upregulation and View blocks and experimental and control groups revealed 11 significant connectivity links in upregulation versus view and between experimental (activity-based neurofeedback) and the sham group as an effect of neurofeedback (Repeated Measures ANOVA for 38 × (38−1)/2 = 703; FDR-corrected for multiple comparison; $q = 0.05$)). Same analysis was done for connectivity-based neurofeedback and in addition to those 11 connections, two more connections including ventral striatum-insula and amygdala-OFC were found. In addition, because of connectivity-based neurofeedback, connectivity (upregulation versus view) of 7 connections out of those 11 links increased significantly in comparison to activity-based neurofeedback. These 7 connections are ventral striatum-insula, amygdala-thalamus, thalamus-insula, thalamus-orbitofrontal cortex, DMPFC-ventral striatum, amygdala-DMPFC, and orbitofrontal cortex – ventrolateral prefrontal cortex. Fig. 3 illustrates the connections established through connectivity-based neurofeedback, A thicker connection between the regions indicates a stronger correlation value.

The EEG frontal asymmetry, represented as LnP(F4)-LnP(F3) in the alpha frequency band and coherence of F4 and F3 electrodes, for upregulation versus view and rest in three experiments is shown in Fig. 4.

Fig. 4. Is a comparison of "LnP(F4)-LnP(F3)" for upregulation versus view and rest of different experiments and shows that "LnP(F4)-LnP(F3)" of upregulation versus view and rest in connectivity-based neurofeedback was higher than activity-based and sham neurofeedback. The greater value of "LnP(F4)-LnP(F3)" in Upregulation compared to View and Rest in connectivity-based neurofeedback is an indicator of increased happiness during emotion regulation, reduction of negative affect and anxiety, and treatment of mental disorders such as PTSD and depression. (Allen and Reznik, 2015; Mennella et al., 2017; Meyer et al., 2015, 2018; Peeters et al., 2014; Reznik and Allen, 2018). The difference

**Table 1**
Percentages of signal change in different regions for Upregulation versus View and Rest for connectivity-based neurofeedback and activity-based neurofeedback (FDR-corrected for multiple comparison; q = 0.05).

| Regions | Activity-based neurofeedback Signal change (%) | | | Connectivity based neurofeedback Signal change (%) | | | Previous studies |
|---|---|---|---|---|---|---|---|
| | UP-View | Up-Rest | t-score (UP-View) | UP-View | UP-Rest | t-score (UP-View) | Sig % UP-Rest |
| Left Amygdala | 0.86 | 0.70 | 4.9 | 1.08 | 0.79 | 5.7 | 0.7 (Young et al. (2014)), 0.3 (Zotev et al. (2016)), 0.3 (Zotev et al. (2016)), 0.1 (Li et al. (2016)), 0.2 (Kim et al. (2015)) |
| Right Amygdala | 0.65 | 0.72 | 3.9 | 0.94 | 0.82 | 6.1 | 0.4 (Young et al. (2014)), 0.3 (Zotev et al. (2016)) |
| Left Insula | 1 | 0.64 | 7.7 | 1.32 | 0.82 | 10.16 | 0.5 (Johnston et al. (2010)), 0.5 (Li et al. (2016)) |
| Right Insula | 0.91 | 0.62 | 6.4 | 1.04 | 0.70 | 10.22 | - |
| Left Anterior Cingulate Cortex | 0.97 | 0.81 | 4.2 | 1.26 | 0.99 | 5.33 | 0.3 (Li et al. (2016)) |
| Right Anterior Cingulate Cortex | 0.64 | 0.38 | 4.4 | 0.75 | 0.8 | 4.6 | - |
| Left Cuneus | 0.45 | 1.56 | 4.5 | - | 1.7 | - | 0.5 (Johnston et al. (2010)) |
| Right Cuneus | 0.4 | 1.9 | 3.9 | - | - | - | - |
| Left Lingual Gyrus | 1.21 | 1.39 | 4.2 | - | - | - | - |
| Left Posterior Cingulate Cortex | 0.49 | 0.33 | 5 | - | 0.19 | - | 0.5 (Johnston et al. (2010)) |
| Left Thalamus | 1.07 | 0.85 | 4.9 | 1.3 | 0.95 | 6.2 | - |
| Right Thalamus | 0.86 | 0.65 | 6.9 | 0.95 | 0.7 | 7 | - |
| Left Caudate | 0.86 | 0.65 | 5.7 | 0.89 | 0.71 | 6.30 | - |
| Right Caudate | 0.74 | 0.49 | 8 | 0.93 | 0.74 | 9.15 | - |
| Left Hippocampus | 0.57 | 0.65 | 4.3 | 0.97 | 0.94 | 5.95 | - |
| Right Hippocampus | 0.44 | 0.59 | 4.2 | 0.62 | 0.79 | 5.1 | - |
| Left Dorsomedial Prefrontal Cortex | 0.85 | 1.02 | 5.1 | 1.26 | 1.36 | 6.35 | - |
| Right Dorsomedial Prefrontal Cortex | 0.37 | 0.81 | 4.1 | 0.95 | 0.94 | 6.71 | - |
| Left Orbitofrontal Cortex | 1.13 | 1.04 | 6 | 1.49 | 1.12 | 6.25 | - |
| Right Orbitofrontal Cortex | 1.12 | 0.81 | 6.5 | 1.3 | 0.9 | 7.63 | - |
| Left Middle Temporal Gyrus | 0.66 | 0.59 | 5.9 | 0.74 | 0.60 | 8.26 | - |
| Right Middle Temporal Gyrus | 0.69 | 0.70 | 5.9 | 0.7 | 0.8 | 7.43 | - |
| Left Ventral Striatum | 1.17 | 0.84 | 6.1 | 1.36 | 0.82 | 9.88 | 0.5 (Johnston et al. (2010)) |
| Right Ventral Striatum | 0.81 | 0.66 | 7.7 | 0.9 | 0.79 | 9.22 | 0.5 (Johnston et al. (2010)) |
| Left Ventrolateral Prefrontal Cortex | 0.67 | 0.81 | 5.6 | 0.82 | 0.84 | 11.38 | 0.5 (Allen and Reznik (2015); Mennella et al. (2017); Meyer et al. (2015), 2018; Peeters et al. (2014); Reznik and Allen (2018)) |
| Right Ventrolateral Prefrontal Cortex | 0.65 | 0.58 | 9.3 | 0.57 | 0.65 | 7.8 | - |
| Left Dorsolateral Prefrontal Cortex | 0.84 | 0.90 | 5.3 | 1 | 1.29 | 7.78 | - |
| Right Dorsolateral Prefrontal Cortex | 0.76 | 0.75 | 6.7 | 0.82 | 1 | 7.27 | - |
| Left Superior Parietal | 0.46 | 0.69 | 4.9 | 0.4 | 0.6 | 4.5 | - |
| Right Superior Parietal | 0.33 | 0.89 | 5.3 | - | 0.7 | - | - |
| Left Inferior Parietal | 0.60 | 0.41 | 5.9 | 0.55 | 0.4 | 5.4 | - |
| Right Inferior Parietal | 1.32 | 0.53 | 4.2 | - | 0.6 | - | - |
| Left SupraMarginal | 0.87 | 0.42 | 8.1 | 0.8 | 0.5 | 5.5 | - |
| Left Postcentral | 0.67 | 0.56 | 6.8 | 0.6 | 0.52 | 5.9 | - |

in "LnP(F4)-LnP(F3)" between Upregulation and View/Rest in the connectivity-based neurofeedback group was significantly different from that of the sham and activity-based neurofeedback groups (two sample *t*-test on mean change of all runs; t(26) $_{Upregulation-View (Connectivity\ nf\ vs.\ Activity\ group)}$ = 2.8, $p_{Upregulation-View (Connectivity\ nf\ vs.\ Activity\ group)}$ = 9 $\times 10^{-3}$; and t(22) $_{Upregulation-View (Connectivity\ nf\ vs.\ Control\ group)}$ = 5.3; $p_{Upregulation-View (Connectivity\ nf\ vs.\ Control\ group)}$ = 2 $\times 10^{-5}$, t(26) $_{Upregulation-Rest (Connectivity\ nf\ vs.\ Activity\ group)}$ = 3.5, $p_{Upregulation-Rest (Connectivity\ nf\ vs.\ Activity\ group)}$ = 1.6 $\times 10^{-3}$; and t(22) $_{Upregulation-Rest (Connectivity\ nf\ vs.\ Control\ group)}$ = 5.9; $p_{Upregulation-Rest (Connectivity\ nf\ vs.\ Control\ group)}$ < 1 $\times 10^{-5}$). In addition, Fig. 4 shows the coherence changes for upregulation versus view and rest for three experimental groups including sham, activity-based neurofeedback, and connectivity-based neurofeedback and it shows significant increases in coherence of F3 and F4 in upregulation versus view and rest for both activity-based and connectivity-based neurofeedback ($p_{Upregulation\ vs.\ View (connectivity-based\ nf)}$ = 1 $\times 10^{-6}$, $p_{Upregulation\ vs.\ Rest (connectivity-based\ nf)}$ = 5.8 $\times 10^{-7}$, $p_{Upregulation\ vs.\ View (activity-based\ nf)}$ = 4 $\times 10^{-5}$, $p_{Upregulation\ vs.\ Rest (activity-based\ nf)}$ = 5 $\times 10^{-5}$).

In our previous study (Dehghani et al., 2021), the coherence analysis of 24 EEG electrodes including AF3/AF4, Fp1/Fp2, F7/F8, F3/F4, FC5/FC6, CP5/CP6, P3/P4, O1/O2, T7/T8, P7/P8, CP1/Cp2, and C3/C4 from all brain regions including frontal, occipital, central, temporal, and parietal was performed on EEG data from both the experimental (EEG frontal asymmetry neurofeedback) and the control groups, and different blocks of the experimental and control groups were compared. The analysis resulted in several pairs of EEG channels with significant coherence changes for Upregulation versus View and Rest blocks and also for experimental versus sham groups (FDR-corrected for multiple comparison, q = 0.05) (Dehghani et al., 2021). These pairs of channels are summarized in Table 2. Additionally in this study, the same coherence analysis was conducted on the experimental group that received connectivity-based neurofeedback. Comparing the coherence of the connectivity-based and activity-based neurofeedback groups revealed increased coherence between several pairs of EEG channels that are highlighted in Table 2.

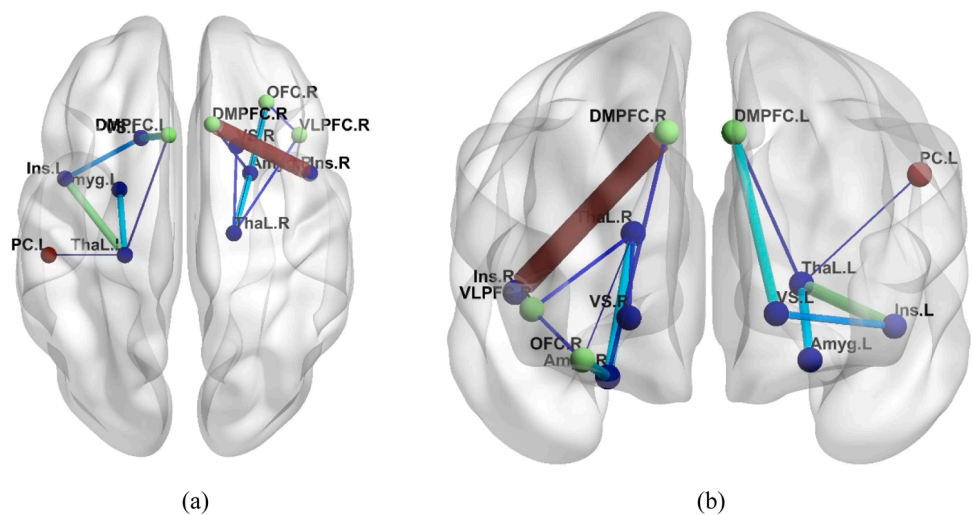

**Fig. 3.** Significant connectivity networks between the Upregulation and View blocks for panel (a) axial, and (b) coronal layout. A thicker connection between the regions indicates a stronger functional connectivity value. L, left; R, right; Amyg, amygdala; ThaL, thalamus; Ins, insula; OFC, orbitofrontal cortex; VS, ventral striatum; DMPFC, dorsomedial prefrontal cortex; VLPFC, ventrolateral prefrontal cortex; PC, postcentral.

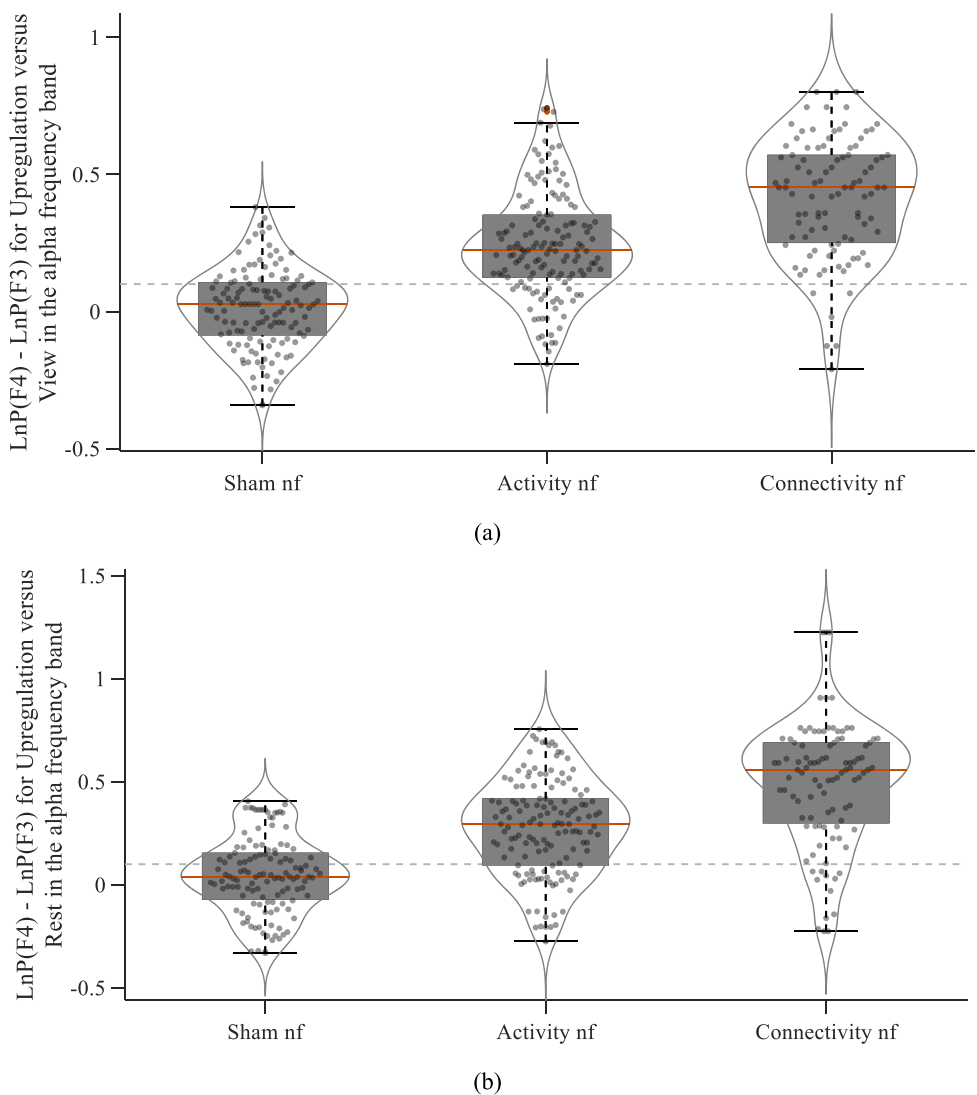

**Fig. 4.** (a)–(b) EEG frontal asymmetry (P(F4)-P(F3)) in the alpha frequency band for upregulation versus view and rest and (c)–(d) Coherence between F3 and F4 in the alpha frequency band for upregulation versus view and rest for three experimental groups including sham, activity-based and connectivity- based neurofeedback.

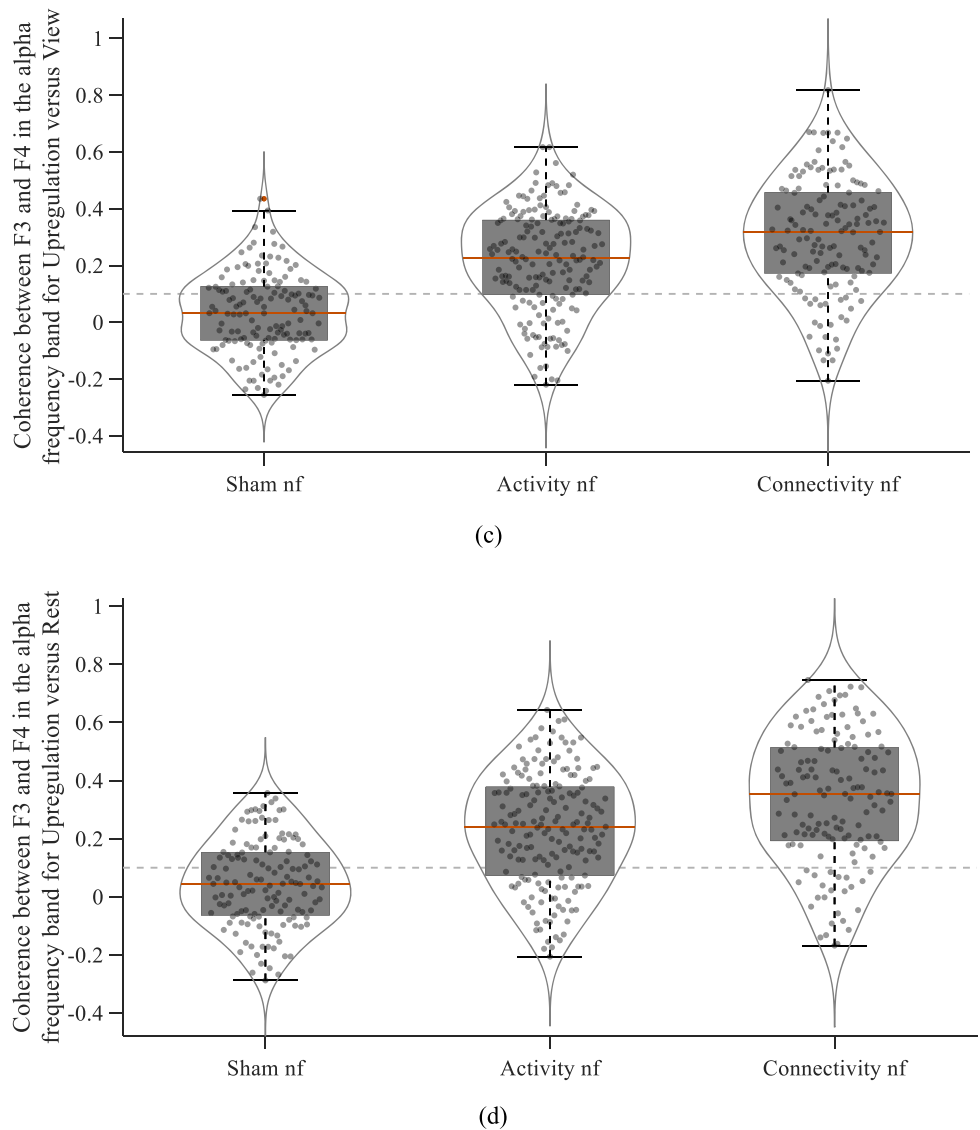

Fig. 4. (*continued*).

**Table 2**
Coherence among EEG electrodes in different frequency bands for Upregulation Versus View and Rest in the connectivity-based neurofeedback group.

| Delta | Theta | Alpha | | Beta |
|---|---|---|---|---|
| Up-View | Up-Rest | Up-View | Up-Rest | Up-Rest |
| **F3-P3** (p= 3.7 × $10^{-4}$) | **CP1-FP2** (p= 5.5 × $10^{-5}$) | **F4-F3** (p= 4 × $10^{-5}$) | F4-F3 (p= 5 × $10^{-5}$) | FP1-FC6 (p= 6 × $10^{-3}$) |
| **F3-FP2** (p= 4.7 × $10^{-4}$) | CP1-F4 (p= 7.4 × $10^{-4}$) | C3-F4 (p= 5.8 × $10^{-5}$) | P7-P4 (p= 6.9 × $10^{-4}$) | C3-CP2 (p= 9.3 × $10^{-5}$) |
| F7-AF4 (p= 6 × $10^{-3}$) | **F8-C3** (p= 5.7 × $10^{-5}$) | F8-C3 (p= 1.8 × $10^{-4}$) | T7-P3 (p= 7.3 × $10^{-4}$) | **CP2-FC6** (p= 7.9 × $10^{-4}$) |
| F7-P3 (p= 7.2 × $10^{-4}$) | F7-O1 (p= 8.3 × $10^{-4}$) | C4-O2 (p= 2.9 × $10^{-4}$) | T8-P3 (p= 3.5 × $10^{-3}$) | |
| | FC5-O1 (p= 2.5 × $10^{-3}$) | **FC5-C4** (p= 8.8 × $10^{-5}$) | **CP6-P3** (p= 4.8 × $10^{-4}$) | |
| | | CP5-T7 (p= 9.5 × $10^{-5}$) | | |

**Table 3**
Psychometric tests of all groups for before and after neurofeedback experiment.

| Psychometric assessment | Activity nf | | Sham nf | | Connectivity nf | |
|---|---|---|---|---|---|---|
| | Pre | Post | Pre | Post | Pre | Post |
| PANAS | 52.2 ± 11.5 | 51.6 ± 8.8 | 54.2 ± 5.9 | 53 ± 4.8 | 50 ± 6.2 | 51.7 ± 5.7 |
| PANAS negative mood states | 20.8 ± 7.2 | 14.1 ± 4.8 | 22.1 ± 6 | 19.3 ± 5 | 21 ± 5.7 | 13.7 ± 3.8 |
| PANAS positive mood states | 31.4 ± 6. | 37.5 ± 6.4 | 32.1 ± 5.7 | 33.7 ± 5.6 | 29 ± 3.2 | 38 ± 5.6 |
| POMS | 24.6 ± 10.9 | 17 ± 6.9 | 27.1 ± 11.3 | 22.8 ± 11.1 | 25.5 ± 8.9 | 18.7 ± 8.8 |
| Total Mood Distribution (TMD) | 7.5 ± 11.5 | -4.7 ± 7.1 | 6.6 ± 11.2 | 3.6 ± 11.7 | 9.8 ± 13.7 | -5.3 ± 10.6 |

The results of the psychometric tests for all experiments are summarized in Table 3 and Fig. 5.

The average scores for the PANAS positive and negative mood states, POMS and TMD in connectivity based neurofeedback group changed significantly from before to after neurofeedback (p-value$_{positive\ mood\ states}$ of PANAS = 5.5 × $10^{-6}$, p-value$_{negative\ mood\ states\ of\ PANAS}$ = 2.7 × $10^{-4}$, p-value$_{POMS}$ = 8.4 × $10^{-5}$ and p-value$_{TMD}$ = 7.6 × $10^{-5}$). The results of the psychometric assessment indicate that connectivity-based neurofeedback leads to greater changes in positive and negative mood states and TMD compared to other experimental groups. Specifically, connectivity-based neurofeedback results in a greater increase in happiness and decrease in sadness when compared to traditional activity-based neurofeedback.

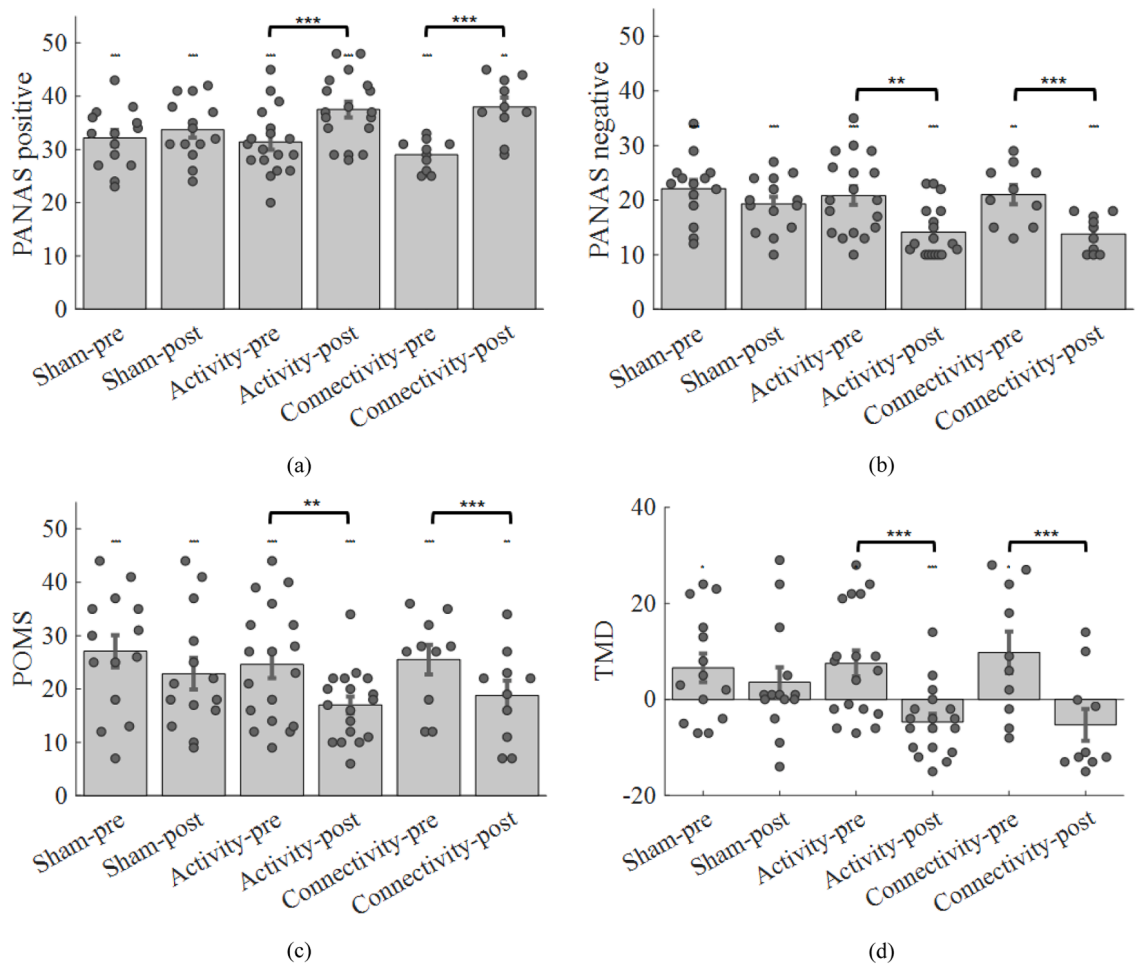

**Fig. 5.** Psychometric assessment before and neurofeedback for all experimental group. (1) PANAS positive (b) PANAS negative, (c) POMS, and (d) TMD. ** = p < 0.01; *** = p < 0.001.

## 5. Discussion

### 5.1. Effectiveness of connectivity-based neurofeedback in emotion regulation

This study aimed to investigate the impact of connectivity-based neurofeedback on emotion regulation and compare it to activity-based and sham neurofeedback. The findings revealed that connectivity-based neurofeedback was more effective in regulating emotions compared to activity-based and sham neurofeedback.

The effectiveness of connectivity-based neurofeedback was demonstrated through various measures, including fMRI BOLD signal changes in emotion regulatory regions, functional connectivity changes, EEG frontal asymmetry, EEG coherence, and psychometric assessments. Notably, the connectivity-based neurofeedback group exhibited higher fMRI BOLD signal changes in emotion regulatory regions, indicating enhanced regulation of emotions. Furthermore, the connectivity-based neurofeedback group showed significant changes in positive and negative emotions, as assessed by the PANAS psychometric assessment. Specifically, the changes in positive emotions in the connectivity-based neurofeedback group were significantly different from those in the activity-based neurofeedback group (p < 0.05), suggesting greater efficacy in regulating positive emotions. Additionally, the change in the total mood distribution (TMD) was higher in the connectivity-based neurofeedback group. This indicates that participants in this group were able to effectively increase/decrease both positive/negative emotions compared to those in the other experimental groups. These findings highlight the superior effectiveness of connectivity-based neurofeedback in emotion regulation. The study provides valuable insights into the potential benefits of targeting connectivity patterns in the brain to enhance emotion regulation processes. Further research in this area could contribute to the development of more targeted and effective interventions for individuals with emotion dysregulation or related conditions.

### 5.2. Neural mechanisms and role of brain regions involved in connectivity-based neurofeedback

The activation map extracted by comparing connectivity-based and activity-based neurofeedback groups in Fig. 2 reveals several brain regions including frontal regions and deep brain region e.g., insula, anterior cingulate cortex, caudate, putamen, thalamus, and temporal and parietal regions. The role of brain regions and connections extracted in this study were discussed in details in previous studies (Dehghani et al., 2022) according to the emotion regulation model (Kohn et al., 2014). Emotion regulation involves various brain regions including subcortical and limbic areas. Previous research consistently demonstrates that specific regions, such as the amygdala, insula, ACC, cuneus, caudate, OFC, VLPFC, ventral striatum, and temporal gyrus are consistently activated during emotion regulation. The limbic system and subcortical regions, such as the hypothalamus, thalamus, amygdala, and hippocampus have significant roles in regulating emotions. Specifically, the amygdala plays a crucial role in emotion regulation and establishes connections with other regions involved in regulating emotions through distinct pathways (Ochsner and Gross, 2008; Phelps, 2004). The thalamus serves as a central hub for relaying sensory information and plays a role in processing a wide range of emotional stimuli (Cerqueira et al., 2008). The insula is responsible for monitoring internal emotional states (Pohl et al., 2013). Increased activity in the thalamus, amygdala, insula, and other regions such as the hippocampus, ACC, and ventral striatum is consistent with their with their functions in regulating emotions and facilitating the recall of positive autobiographical memories (Bush et al., 2000; Suardi et al., 2016).

The lingual gyrus and cuneus in the occipital lobe are involved in visual processing and have been implicated in working memory, cognitive functions, visual processing, and memory retrieval in previous studies. Therefore, their increased activity in this study is likely related to the visual processing of positive autobiographical memory images (Burianova et al., 2010; Deak et al., 2017; Vrticka et al., 2013).

The prefrontal and frontal cortex regions, which encompass the OFC, VLPFC, and DLPFC, play vital roles in various cognitive processes such as emotion generation and regulation, self-monitoring, working memory, recalling autobiographical memories, decision making, and processing positive emotional stimuli (Schutter and van Honk, 2006; Svoboda et al., 2006). These regions establish reciprocal connections with other areas involved in emotion regulation including amygdala and ACC. Consequently, the heightened activity observed in the prefrontal and frontal cortex during this study corresponds to their involvement in the regulation of emotions.

The increased activity in parietal regions, comprising the superior parietal, inferior parietal, supramarginal, and postcentral areas can be ascribed to their engagement in directing attention towards positive autobiographical memory images and the integration of sensory and behavioral information (Aday et al., 2017; Bullier, 2001).

*5.3. Effects of connectivity-based neurofeedback on functional connectivity*

Functional connectivity analysis in the connectivity-based neurofeedback group resulted in the identification of two additional connections: ventral striatum-insula and amygdala-OFC. These findings indicate that connectivity-based neurofeedback had a broader effect on functional connectivity, involving additional connections beyond those observed in activity-based neurofeedback.

The increased functional connectivity resulting from connectivity-based neurofeedback suggests greater involvement of key regions in emotion regulation, such as the thalamus, and prefrontal regions, leading to improved emotion regulation as evidenced by changes in EEG/fMRI signals and psychometric evaluations, as per three steps of emotion regulation described in (Kohn et al., 2014).

According to the emotion regulation model proposed by (Kohn et al., 2014), emotion regulation consists of three steps. Firstly, the stimulus is appraised, and subcortical regions such as the amygdala, basal ganglia, and ventral striatum are involved in generating emotions. Secondly, the need for regulation is detected and signaled by the ventrolateral prefrontal cortex (VLPFC) and insula. Finally, regulation takes place, leading to a change in emotional state. The thalamus plays a role in directing sensory information to different cortical and subcortical regions. Previous study has shown that increased functional connectivity between the amygdala and thalamus, as well as between the ventral striatum and thalamus, is associated with enhanced emotion generation and reduced severity of major depressive disorder (Tang et al., 2018).

The increased functional connectivity between the thalamus and insula indicates the involvement of the insula in signaling the need for regulation and its role in the later stages of emotion regulation. This connectivity has been observed during meditation and is similar to the effects of neurofeedback (Jang et al., 2018). Similarly, the increased functional connectivity between the thalamus and VLPFC can be interpreted in a similar manner. The OFC receives input and sensory information from various brain regions and the increased connectivity between the OFC and thalamus is due to the relay of sensory information. There are indirect connections between the subcortical and prefrontal regions involved in emotion regulation. The functional connectivity between different prefrontal regions, limbic/paralimbic regions, and the thalamus is supported by anatomical connections.

Higher co-activation between the limbic system (amygdala, insula, ventral striatum, and thalamus) and prefrontal/frontal cortex (dorsomedial prefrontal cortex (DMPFC, OFC, and VLPFC) is associated with effective emotion regulation and reduced negative affect and anxiety. However, individuals with major depressive disorder exhibit lower functional connectivity between the thalamus/caudate and OFC (Banks et al., 2007; Cheng et al., 2016; Dehghani et al., 2022; Kong et al., 2018). The increased functional connectivity between the thalamus and the postcentral gyrus (a part of the parietal lobe) is related to the parietal lobe's role in receiving and processing sensory information relayed by the thalamus.

Moreover, the study found that connectivity-based neurofeedback had a significant impact on the connectivity (upregulation versus view) of seven connections out of the initial 11 significant links. These seven connections are ventral striatum-insula, amygdala-thalamus, thalamus-insula, thalamus-orbitofrontal cortex, DMPFC-ventral striatum, amygdala-DMPFC, and orbitofrontal cortex-ventrolateral prefrontal cortex.

*5.4. Implications and limitations*

These findings are important in understanding the effects of different types of neurofeedback on functional connectivity in the brain. The identification of specific connections influenced by neurofeedback can provide insights into the underlying neural mechanisms of emotion regulation and potentially guide the development of targeted interventions.

Overall, the results of this study provide novel insights into the effects of connectivity-based neurofeedback on functional connectivity in the brain. The identification of specific connections influenced by neurofeedback contributes to our understanding of the neural mechanisms underlying emotion regulation. These findings have the potential to inform the development of targeted interventions for individuals with emotion dysregulation or related disorders.

A limitation of this study is utilizing of solely F3 and F4 EEG channels for the connectivity-based neurofeedback. The deliberate focus on these two channels was based on our previous research (Dehghani et al., 2021) and was intended to facilitate a direct comparison with the widely established EEG frontal asymmetry approach. However, the choice to analyze only two EEG channels may limit the generalizability of our findings. Future research should explore a more comprehensive analysis involving a wider array of EEG channels, ideally spanning various brain regions. This approach could provide a more comprehensive understanding of the neural mechanisms underlying emotion regulation and the potential effects of connectivity-based neurofeedback. By incorporating data from a broader range of EEG channels, we could better capture the intricate dynamics of brain connectivity and their relation to emotional states.

The other limitation of this study was its exclusive focus on males due to challenges encountered in enrolling female participants. Consequently, the generalizability of the findings may be limited, considering previous reports highlighting gender-specific differences in emotion regulation (A. K. Mak et al., 2009; Mak et al., 2009; McRae et al., 2008). To address this limitation, future studies incorporate female subjects into the same emotion regulation paradigm to provide a more comprehensive understanding of the topic. Due to the significant expenses, time requirements, and practical constraints associated with simultaneous EEG and fMRI recordings, such studies often have small sample sizes (Berman et al., 2013; Cury et al., 2019; Ebrahimzadeh et al., 2021b, 2021a; Kinreich et al., 2012; Lioi et al., 2020; Perronnet et al., 2017; Zich et al., 2015; Zotev et al., 2016, 2014). Nonetheless, when comparing the participant count in our study (n = 42) to that of a recent literature review on emotion regulation-based fMRI neurofeedback (Linhartová et al., 2019), our sample size matches or surpasses that of 53 out of the 55 studies reviewed.

## 6. Conclusion

In this study, a novel EEG neurofeedback technique based on connectivity was used, combined with positive autobiographical memories and simultaneous fMRI to modify brain function. Previous studies have demonstrated the effectiveness of both traditional activity-based neurofeedback and the approach used in this study for altering brain activity during emotion regulation (Dehghani et al., 2022, 2021, 2020; Mosayebi et al., 2022). By utilizing connectivity-based neurofeedback, significant increases were observed in brain regions such as the amygdala, insula, thalamus, and frontal/prefrontal regions, as indicated in Table 1. There was also an enhancement in the interaction and connectivity among multiple brain regions, including the prefrontal and limbic regions. Moreover, there were positive changes in EEG frontal asymmetry in the alpha frequency band, which is a biomarker associated with the treatment of PTSD, depression, and anxiety. These results were further supported by increased coherence between EEG channels and improvements in happiness and reductions in sadness according to psychometric assessments.

The findings demonstrate that connectivity-based neurofeedback surpasses activity-based neurofeedback in enhancing emotion regulation and providing greater voluntary control over brain and mental functions. Compared to activity-based neurofeedback, utilizing connectivity-based neurofeedback in the treatment of mental disorders, particularly major depressive disorder (MDD) and post-traumatic stress disorder (PTSD), could yield more effective modulation of behavior and cognition. The proposed paradigm and connectivity neurofeedback method have the potential to be applied in future studies aimed at developing improved treatments and interventions for mental disorders, focusing on the modulation of connectivity patterns in the brain.

These results emphasize the superior efficacy of connectivity-based neurofeedback in facilitating emotion regulation. The study offers valuable insights into the potential advantages of targeting connectivity patterns in the brain to enhance processes related to emotion regulation. Further research in this area could contribute to the development of more precise and effective interventions for individuals experiencing emotion dysregulation or related conditions.

## Author statement

We confirm that the manuscript has been read and approved by all named authors and that there are no other persons who satisfied the criteria for authorship but are not listed. We further confirm that the order of authors listed in the manuscript has been approved by all of us.

## Code and data availability

The code and data is available on request from the authors.

## Declaration of Competing Interest

The authors declare that they have no known competing financial interests or personal relationships that could have appeared to influence the work reported in this paper.

## Data availability

Data will be made available on request.